# Spin density wave as a superposition of two magnetic states of opposite chirality and its implications

Elijah E. Gordon[a], Shahab Derakhshan[b], Corey M. Thompson[c], and Myung-Hwan Whangbo[a,d,e,*]

[a] Department of Chemistry, North Carolina State University, Raleigh, NC 27695-8204, USA

[b] Department of Chemistry and Biochemistry, California State University Long Beach, Long Beach, California 90840, USA

[c] Department of Chemistry, Purdue University, West Lafayette, Indiana 47907-2084, USA

[d] Group SDeng, State Key Laboratory of Structural Chemistry, Fujian Institute of Research on the Structure of Matter (FJIRSM), Chinese Academy of Sciences (CAS), Fuzhou, China, 350002

[e] State Key Laboratory of Crystal Materials, Shandong University, Jinan, China, 250100

**ABSTRACT:** A magnetic solid with weak spin frustration tends to adopt a noncollinear magnetic structure such as cycloidal structure below a certain temperature and a spin density wave (SDW) slightly above this temperature. The causes for these observations were explored by studying the magnetic structures of $BaYFeO_4$, which undergoes an SDW and a cycloidal phase transition below 48 and 36 K, respectively, in terms of density functional theory calculations. We show that an SDW structure arises from a superposition of two magnetic states of opposite chirality, an SDW state precedes a chiral magnetic state due to the lattice relaxation, and whether an SDW is transversal or longitudinal is governed by the magnetic anisotropy of magnetic ions.

A magnetic system composed of identical magnetic ions tends to adopt a noncollinear magnetic superstructure such as a cycloid (**Fig. 1a**) below a certain temperature when its spin exchange interactions are weakly spin frustrated,[1-6] while a magnetic system with strong spin frustration does not undergo a log-range magnetic ordering.[7] It is often observed that a spin density wave (SDW) state (**Fig. 1b,c**) occurs slightly above the onset temperature of the cycloid.[8-13] In a cycloidal structure, all moments of the ions are the same in magnitude although they differ in orientation (**Fig. 1a**). In an SDW (**Fig. 1b,c**), the moments of the magnetic ions change their magnitudes depending on their positions in the crystal lattice, i.e., they vary sinusoidally along the propagation direction of the SDW and may even vanish. The latter is apparently unphysical for any magnetic system consisting of identical magnetic ions although, mathematically, they correctly describe the observed magnetic structure. So far, this apparently puzzling observation has not been explained to the best of our knowledge, nor has been why a cycloidal state is preceded by a SDW. Furthermore, a SDW can be transversal (**Fig. 1b**) as found for $BaYFeO_4$[8] or longitudinal (**Fig. 1c**) as found for $TbMnO_3$.[9,10] What controls whether a SDW is transversal or longitudinal is not well understood. In the present work, we explore the answers to these questions by studying the magnetic oxide $BaYFeO_4$ (BYFO), which undergoes a SDW and cycloidal magnetic phase transitions below 48 and 36 K, respectively.[8,14]

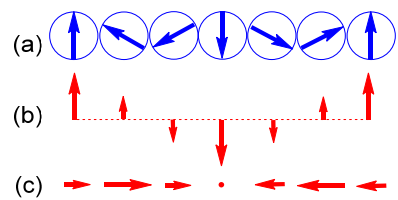

Figure 1. Magnetic superstructures typically observed for a magnetic system with weak spin frustration: (a) a cycloid. (b) a transverse SDW. (c) a longitudinal SDW. For the sake of simplicity, the SDWs and cycloids illustrated here are commensurate ones. Our discussion is equally valid for incommensurate SDWs and incommensurate cycloids.

BYFO is made up of corner-sharing $FeO_6$ octahedra and $FeO_5$ square pyramids, both containing high-spin $Fe^{3+}$ ($d^5$, S = 5/2) ions.[8,14] These polyhedra share their corners to form $Fe_4O_{18}$ tetramer units (**Fig. 2a**), and these tetramers share their corners to form folded-ladder (FL) chains of formula $(FeO_4)_4$ running along the b-direction (**Fig. 2b**). Such FL-chains are packed in BYFO with two types of orientations, A and B (**Fig. 2c**). The axial Fe-O bonds of the $FeO_5$ square pyramids are aligned along the (a – c) and (a + c) directions in type A and B orientations, respectively. The magnetic susceptibility of BYFO hints two antiferromagnetic (AFM) transitions at 36 and 48 K.[14] The neutron powder diffraction study revealed that the two magnetic transitions at 36 and 48 K arise from a long-range magnetic order.[7] The magnetic structure below 48 K is best described by a transversal SDW (**Fig. 1b**) with propagation vector k = (0, 0, 1/3) and moments along the b-axis, and the magnetic structure below 36 K by a cycloid (**Fig. 1a**) with propagation vector k

= (0, 0, 0.358) and moments lying in the bc-plane. The cycloidal structure below 36 K lacks inversion symmetry, so BYFO exhibits ferroelectric polarization induced by a magnetic order.[15] TbMnO$_3$ adopts a longitudinal SDW (**Fig. 1c**) with moments in the b-direction below 42 K and a cycloidal (**Fig. 1a**) magnetic structure with moments in the bc-plane below 28 K.[9,10] In what follows, we examine the nature of spin frustration in BYFO leading to its SDW and cycloidal states as well as under what condition they occur by evaluating six intrachain and four interchain spin exchanges (**Fig. 2b,c**) and determining the magnetic anisotropy of the Fe$^{3+}$ ions on the basis of density functional theory (DFT) calculations. Subsequently, we explore important implications of our finding.

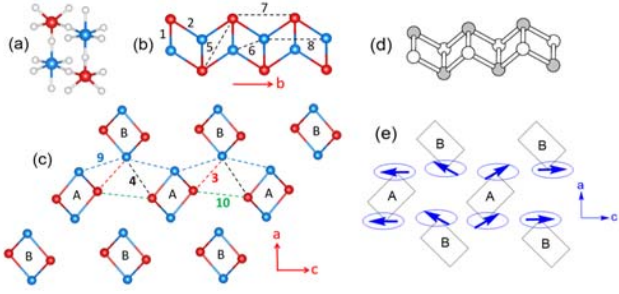

FIGURE 2. Schematic description of the crystal structure, the 10 spin exchange paths and a schematic representation of a cycloidal magnetic structure of BaYFeO$_4$: (a) A Fe$_4$O$_{18}$ tetramer unit made up of corner-sharing FeO$_6$ octahedra and FeO$_5$ square pyramids. (b) A folded-ladder chain of formula Fe$_4$O$_{16}$ running along the b-direction. This chain results from Fe$_4$O$_{18}$ tetramers by sharing their oxygen corners. For simplicity, each Fe$_4$O$_{18}$ tetramer is represented by a Fe$_4$ tetramer. (c) A projection view of how the folded-ladder chains are packed in BaYFeO$_4$, in which the folded-ladder chains occur in two types of orientations A and B. The numbers 1 – 10 in (b) and (c) represent the spin exchange paths J$_1$ – J$_{10}$, respectively. (d) In each folded-ladder chain, the nearest-neighbor Fe$^{3+}$-spins are antiferromagnetically coupled. The latter is indicated by using shaded and unshaded spheres. (e) A view of how the spin moment of each Fe$^{3+}$ ion rotates in the bc-plane. The rotational plane at each spin site is represented by an ellipse. For simplicity, only the Fe$^{3+}$-spins forming the interchain exchange paths J$_9$ are shown. The angle between two adjacent spin sites in each exchange path J$_9$ along the c-direction is θ.

We carry out spin-polarized DFT calculations using the Vienna ab Initio Simulation Package,[16] the projector augmented wave method,[17] the PBE exchange-correlation functionals,[18] the plane wave cutoff energy of 500 eV, a set of 2×2×1 k-points for sampling the irreducible Brillouin zone, and the threshold of 10$^{-6}$ eV for self-consistent-field energy convergence. The electron correlation associated with the Fe 3$d$ states were taken into consideration by performing the DFT+U calculations[19] with the effective on-site repulsion $U^{eff} = U - J$ (= 4, 5 eV) on Fe. The preferred spin directions of the Fe$^{3+}$ ions were determined by DFT+U+SOC calculations.[20]

To explore the nature of the noncollinear magnetic structures of BYFO, we examine the six intrachain spin exchanges J$_1$, J$_2$, J$_5$, J$_6$, J$_7$ and J$_8$ (**Fig. 2b**) as well as the four interchain spin exchanges J$_3$, J$_4$, J$_9$ and J$_{10}$ (**Fig. 2c**). The geometrical parameters associated with these exchange paths are summarized in **Table S1**. To extract the values of these spin exchanges, we construct 11 ordered spin states of BYFO (**Fig. S1**) using a (a, 2b, 2c) supercell, which has 32 formula units (FUs). The magnetic energy spectrum of BYFO can be described in terms of the Heisenberg spin Hamiltonian,

$$\hat{H}_{spin} = \sum_{i>j} J_{ij} \vec{S}_i \cdot \vec{S}_j \qquad (1)$$

where J$_{ij}$ = J$_1$ – J$_{10}$. The total spin exchange energies of these states per (a, 2b, 2c) supercell can be expressed as

$$E = \sum_{i=1}^{10} n_i J_i S^2 \qquad (2)$$

where S refers to the spin of the high-spin Fe$^{3+}$ ion (i.e., S = 5/2). The values of n$_i$ (i = 1 – 10) found for the 11 ordered spin states are listed in **Table S2**. The relative energies of these states determined by DFT+U calculations are summarized in **Table S3**. We obtain the values of J$_1$ – J$_{10}$ by energy-mapping analysis,[1] in which the relative energies from the DFT+U calculations are mapped onto the corresponding energies expected from the spin exchange energies (Eq. 2). The values of J$_1$ – J$_{10}$ obtained for the 280, 38 and 6 K structures[7] of BYFO by DFT+U calculations with U$^{eff}$ = 4 and 5 eV are summarized in **Table S4** and **S5**, respectively.

For the 280, 38 and 6 K crystal structures, all spin exchanges (except for J$_9$ at 6 K) are AFM, and the intrachain spin exchanges J$_1$ and J$_2$ dominate over all other intrachain spin exchanges as well as all the interchain spin exchanges. Due to the strong spin exchanges J$_1$ and J$_2$, the NN spins in each FL-chain are strongly coupled antiferromagnetically (**Fig. 2d**) and their collinear spin arrangement should be most stable. This explains the collinear AFM arrangement within each FL-chain observed for the cycloidal magnetic structure at 6 K. We now examine the magnetic anisotropy of the Fe$^{3+}$ (S = 5/2, L = 0) ions of BYFO. Formally, L = 0 for the high-spin Fe$^{3+}$ ions, so the effect of SOC is weak. Thus, in determining the magnetic anisotropy of the Fe$^{3+}$ ions at low temperature, we employ a commensurate, low-energy AFM state of BYFO (e.g., **Fig. S2**) predicted by the spin exchange constants obtained from the DFT+U calculations. Using this AFM state, we perform DFT+U+SOC calculations with three different spin orientations, i.e., parallel to the a-, b- and c-directions. These calculations show that the ||b spin orientation is slightly more stable than the ||c spin orientation, while these two orientations are substantially more stable than the ||a spin orientation (**Table S6** and **S7**). Thus, the spins of the Fe$^{3+}$ ions prefer to lie in the bc-plane, which agrees with the observation that the Fe$^{3+}$ moments lie in the bc-plane in the cycloidal magnetic structure at 6 K.[8]

The cycloidal magnetic structure is determined by the noncollinear spin arrangement in the exchange paths J$_9$ (**Fig. 2c**), which run along the c-direction. As depicted in **Fig. 2e**, the spins in each zigzag chain of J$_9$ paths rotate by the angle of θ in the bc-plane as they move from one spin site to another along the c-direction. In the cycloidal structure of BYFO at 6 K with propagation vector k = (0, 0, 0.358), which is equivalent to 2θ = 128.8° (i.e., θ = 64.4°).[7] This successive spin-rotation is related to the spin frustration between FL-chains, which is brought about by the interchain spin exchanges $J_3$, $J_4$, $J_9$ and $J_{10}$ (**Fig. 2c**) together with the intrachain exchanges $J_1$ and $J_2$. These exchanges form ($J_3$, $J_4$, $J_{10}$), ($J_1$, $J_3$, $J_9$) and ($J_2$, $J_4$, $J_9$) triangles (see **Fig. S3** for more details). According to the packing pattern of the FL-chains (**Fig. 2c**), one FL-chain of a particular arrangement, say, A, interacts with four FL-chains of arrangement B and with two FL-chains of arrangement A. Thus, the total spin exchange energy E(θ) that one FL-chains generates (per four FUs) is written as

$$E(\theta) = 4[(2J_3 + J_4 - 2J_9)\cos\theta - J_{10}\cos 2\theta] \quad (3)$$

The angle $\theta_{min}$ minimizing this energy is obtained by requiring that dE(θ)/dθ = 0, which leads to the expression

$$\theta_{min} = \cos^{-1}\left(\frac{2J_3 + J_4 - 2J_9}{4J_{10}}\right) \quad (4)$$

This expression predicts the $\theta_{min}$ values of 85.7°, 85.1° and 67.3° for the 280, 38 and 6 K structures, respectively, using the $J_3$, $J_4$, $J_9$ and $J_{10}$ values of the $U^{eff}$ = 4 eV calculations. The $\theta_{min}$ value of 67.3° predicted for the 6 K structure is quite close to the experimental value of 64.4°.

Note that a cycloidal magnetic state (**Fig. 1a**) is chiral, and a cycloid of one chirality is identical in energy with that of its opposite chirality as long as the lattice of the magnetic ions remains the same. Then, the two cycloidal states of opposite chirality depicted in **Fig 3a,b** should be equally populated to give rise to a transversal SDW of **Fig. 1b**, because the components of the moments along the propagation direction are canceled out while those perpendicular to the propagation direction are reinforced with their magnitudes varying sinusoidally along the propagation direction. The opposite happens when the cycloidal sates of opposite chirality depicted in **Fig 4a,b** are equally populated, leading to a longitudinal SDW of **Fig. 1c**. In short, when equally populated, two cycloidal states of opposite chirality generate a SDW magnetic structure. The apparently puzzling picture of a SDW, namely, that the moments of the magnetic ions change their magnitudes depending upon their positions in the lattice, is a direct consequence of equally populating the two cycloidal states of opposite chirality. Helical magnetic states are also chiral, so it is possible to have a SDW as a superposition of two helical states of opposite chirality.

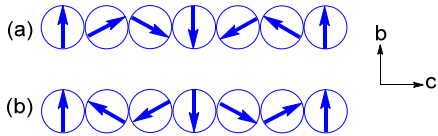

FIGURE 3. Two cycloids of opposite chirality with moments in the bc-plane, leading to a transversal SDW (Fig. 1b) propagating along the c direction.

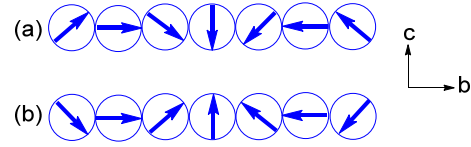

FIGURE 4. Two cycloids of opposite chirality with moments in the bc-plane, leading to a longitudinal SDW shown (Fig. 1c) propagating along the b direction.

On lowering the temperature, a magnetic solid with a SDW state will adopt a cycloidal magnetic structure because its crystal lattice will be relaxed to become energetically more favorable to one of the two cycloidal states so the two cycloidal states become nondegenerate. On lowering the temperature, therefore, a SDW state should be followed by a cycloidal state, as found experimentally.[8-13] In $NaFeSi_2O_6$, a SDW is followed by a helical magnetic state.[21] We note that, in reproducing the entire experimental phase diagram of $RMnO_3$ (R = rare earth), it is necessary to include the spin-phonon coupling,[22,23] which is essential for a crystal lattice to relax and energetically favor one of the two cycloidal states. In this phase diagram study based on a model Hamiltonian, neither the origin of the SDW nor its implications was examined.

Finally, we consider why the SDW of BYFO is transversal (**Fig. 1b**) while that of $TbMnO_3$ is longitudinal (**Fig. 1c**). In BYFO, the moment orientation along the ||b direction is slightly more stable than that along the ||c direction (**Table S6**, **S7**). In the SDW of BYFO, the two cycloids (**Fig. 3a**, **b**) give rise to the nonzero moments perpendicular to the SDW propagation direction (**Fig. 1b**). The preferred orientation of the $Mn^{3+}$ spin in $TbMnO_3$ is contained in the ab-plane and is aligned approximately along the b-direction (**Fig. S4**).[2,12] Due to this magnetic anisotropy plus the ferromagnetic spin exchanges between the nearest-neighbor $Mn^{3+}$ spins,[2] the $Mn^{3+}$ spins prefer to align along the b-axis. In the SDW of $TbMnO_3$, the two cycloids (**Fig. 4a**, **b**) lead to the nonzero moments along the SDW propagation direction (**Fig. 1c**).

In summary, a SDW state is a superposition of two cycloidal states of opposite chirality. A spin density wave state precedes a cycloidal magnetic state, and the magnetic anisotropy of magnetic ions determines whether a SDW is transversal or longitudinal.

**Corresponding Author**

* mike_whangbo@ncsu.edu.

**Funding Sources**

SD is grateful for the financial support from NSF-DMR-RUI Award #1601811.

**ACKNOWLEDGMENT**

MHW would like to thank V. Ovidiu Garlea and Reinhard K. Kremer for invaluable discussions, and the High Performance Computing Services of NCSU for computing resources.

## REFERENCES

(1) Xiang, H. J.; Lee, C.; Koo, H.-J.; Gong, X. G.; Whangbo, M.-H. Magnetic properties and energy-mapping analysis, *Dalton Trans.*, **2013**, *42*, 823-853.
(2) Xiang, H. J.; Wei, S.-H.; Whangbo, M.-H.; Da Silva, J. L. F. Spin-Orbit Coupling and Ion Displacements in Multiferroic TbMnO$_3$, *Phys. Rev. Lett.* **2008**, *101*, 037209.
(3) Kan, E. J.; Xiang, H. J.; Zhang, Y.; Lee, C.; Whangbo, M.-H. Density-functional analysis of spin exchange and ferroelectric polarization in AgCrO$_2$, *Phys. Rev. B*, **2009**, *80*, 104417.
(4) Lee, C.; Kan, E. J.; Xiang, H. J.; Whangbo, M.-H. Theoretical Investigation of the Magnetic Structure and Ferroelectric Polarization of the Multiferroic Langasite Ba$_3$NbFe$_3$Si$_2$O$_{14}$, *Chem. Mater.* **2010**, *22*, 5290-5295.
(5) Lu, X. Z.; Whangbo, M.-H.; Dong, S.; Gong, X. G.; Xiang, H. J. Giant Ferroelectric Polarization of CaMn$_7$O$_{12}$ Induced by a Combined Effect of Dzyaloshinskii-Moriya Interaction and Exchange Striction, *Phys. Rev. Lett.* **2012**, *108*, 187204.
(6) Yang, J. H.; Li, Z. L.; Lu, X. Z.; Whangbo, M.-H.; Wei, S.-H.; Gong, X. G.; Xiang, H. J. Strong Dzyaloshinskii-Moriya Interaction and Origin of Ferroelectricity in Cu$_2$OSeO$_3$, *Phys. Rev. Lett.* **2012**, *109*, 107203.
(7) Greedan, J. E. Geometrically frustrated magnetic materials. *J. Mater. Chem.* **2001**, *11*, 37.
(8) Thompson, C. M.; Greedan, J. E.; Garlea, V. O.; Flacau, R.; Tan, M.; Nguyen, P.-H. T.; Wrobel, F.; Derakhshan, S. Partial Spin Ordering and Complex Magnetic Structure in BaYFeO$_4$: A Neutron Diffraction and High Temperature Susceptibility Study, *Inorg. Chem.* **2014**, *53*, 1122-1127.
(9) Senff, D.; Link, P.; Hradil, K.; Hiess, A.; Regnault, L. P.; Sidis, Y.; Aliouane, N.; Argyriou, D. N.; Braden, M. Magnetic Excitations in Multiferroic TbMnO$_3$: Evidence for a Hybridized Soft Mode, *Phys. Rev. Lett.* **2007**, *98*, 137206.
(10) Senff, D.; Aliouane, N.; Argyriou, D. N.; Hiess, A.; Regnault, L. P.; Link, L.; Hradil, K.; Sidis, Y.; Braden, M., Magnetic excitations in a cycloidal magnet: the magnon spectrum of multiferroic TbMnO$_3$, *J. Phys.: Condens. Matter* **2008**, *20*, 434212.
(11) Prokhnenko, O.; Feyerherm, R.; Dudzik, E.; Landsgesell, S.; Aliouane, N.; Chapon, L.; Argyriou, D. Enhanced Ferroelectric Polarization by Induced Dy Spin Order in Multiferroic DyMnO$_3$, *Phys. Rev. Lett.* **2007**, *98*, 057206.
(12) Sobolev, A.; Rusakov, V.; Moskvin, A.; Gapochka, A.; Belik, A.; Glazkova, I.; Akulenko, A.; Demazeau, G.; Presniakov, I. $^{57}$Fe Mössbauer study of unusual magnetic structure of multiferroic 3R-AgFeO$_2$, J. Phys.: Condens. Matter, **2017**, *29*, 275803.
(13) Terada, N.; Khalyavin, D. D.; Manuel, P.; Tsujimoto, Y.; Knight, K.; Radaelli, P. G.; Suzuki, H. S.; Kitazawa, H. Spiral-Spin-Driven Ferroelectricity in a Multiferroic Delafossite AgFeO$_2$, *Phys. Rev. Lett.* **2012**, *109*, 097203.
(14) Wrobel, F.; Kemei, M. C.; Derakhshan, S. Antiferromagnetic Spin Correlations Between Corner-Shared [FeO$_5$]$^{7-}$ and [FeO$_6$]$^{9-}$ Units, in the Novel Iron-Based Compound: BaYFeO$_4$, *Inorg. Chem.* **2013**, *52*, 2671-2677.
(15) Cong, J.-Z.; Shen, S.-P.; Chai, Y.-S.; Yan, L.-Q.; Shang, D.-S.; Wang, S.-G.; Sun, Y. Spin-driven multiferroics in BaYFeO$_4$, *J. Appl. Phys.* **2015**, *117*, 174102.
(16) Kresse, G.; Furthmüller, J. Efficiency of ab-initio total energy calculations for metals and semiconductors using a plane-wave basis set, *Comput. Mater. Sci.* **1996**, *6*, 15-50.
(17) Kresse, G.; Joubert, D. From ultrasoft pseudopotentials to the projector augmented-wave method, *Phys. Rev. B* **1999**, *59*, 1758-1775.
(18) Perdew, J. P.; Burke, K.; Ernzerhof, M. Generalized Gradient Approximation Made Simple, *Phys. Rev. Lett.*, **1996**, *77*, 3865.
(19) Dudarev, S. L.; Botton, G. A.; Savrasov, S. Y.; Humphreys, C. J.; Sutton, A. P. Electron-energy-loss spectra and the structural stability of nickel oxide: An LSDA+U study, *Phys. Rev. B* **1998**, *57*, 1505.
(20) Kuneš, K.; Novák, P.; Schmid, R.; Blaha, P.; Schwarz, K. Electronic structure of fcc Th: Spin-orbit calculation with 6p1/2 local orbital extension, *Phys. Rev. Lett.,* **2001**, *64*, 153102.
(21) Baum, M. M., Neutron-Scattering Studies on Chiral Multiferroics, Dissertation, Universität zu Köln (2013).
(22) Mochizuki, M.; Furukawa, N.; Nagaosa, N. Theory of spin-phonon coupling in multiferroic manganese perovskites RMnO$_3$. *Phys. Rev. B* **2011**, *84*, 144409.
(23) Mochizuki, M.; Furukawa, N.; Nagaosa, N. Spin Model of Magnetostrictions in Multiferroic Mn Perovskites. *Phys. Rev. Lett.* **2010**, *105*, 037205; Erratum, *Phys. Rev. Lett.* **2011**, *106*, 119901.

SYNOPSIS TOC

A magnetic solid with weak spin frustration tends to adopt a cycloidal structure below a certain temperature and a spin density wave (SDW) slightly above this temperature. Our analysis shows that a SDW arises from a superposition of two cycloids of opposite chirality. Implications of this finding were explored.

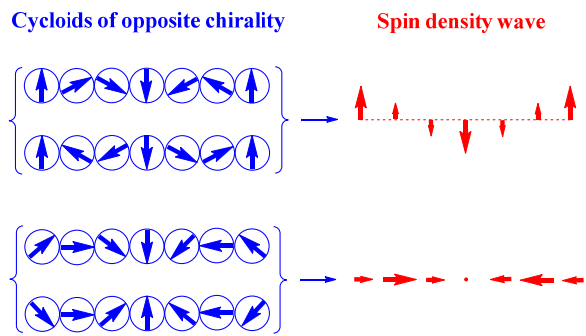

Supporting Information

for

**Spin density wave as a superposition of two magnetic states of opposite chirality and its implications**

Elijah E. Gordon[a], Shahab Derakhshan[b], Corey M. Thompson[c] and Myung-Hwan Whangbo*,[a,d,e]

Table S1. The geometrical parameters associated with the 10 spin exchange paths $J_1 - J_{10}$ of BaYFeO$_4$ in the 280, 38 and 6 K structures.

| Path | Nature [a] | $Fe^{3+}\cdots Fe^{3+}$ distance (Å) | | |
|---|---|---|---|---|
| | | 280 K | 38 K | 6 K |
| $J_1$ | Intra, SPYD-OCT | 3.789 | 3.779 | 3.771 |
| $J_2$ | Intra, SPYD-OCT | 4.055 | 4.044 | 4.024 |
| $J_3$ | Inter, SPYD-OCT | 5.948 | 5.932 | 5.990 |
| $J_4$ | Inter, SPYD-OCT | 4.830 | 4.817 | 4.859 |
| $J_5$ | Intra, SPYD-SPYD | 5.329 | 5.315 | 5.296 |
| $J_6$ | Intra, OCT-OCT | 5.762 | 5.746 | 5.726 |
| $J_7$ | Intra, SPYD-SPYD | 5.695 | 5.678 | 5.698 |
| $J_8$ | Intra, OCT-OCT | 5.695 | 5.678 | 5.698 |
| $J_9$ | Inter, OCT-OCT | 6.072 | 6.056 | 6.090 |
| $J_{10}$ | Inter, SPYD-SPYD | 6.526 | 6.509 | 6.579 |

[a] SPYD and OCT refer to the FeO$_5$ square pyramidal and the octahedral FeO$_6$, respectively.

Table S2. The values of $n_1 - n_{10}$ needed to specify the total spin exchange energies per 32 FUs in Eq. 2 for the 11 ordered spin states of BaYFeO$_4$

|      | $n_1$ | $n_2$ | $n_3$ | $n_4$ | $n_5$ | $n_6$ | $n_7$ | $n_8$ | $n_9$ | $n_{10}$ |
|------|-----|-----|-----|-----|-----|-----|-----|-----|-----|------|
| FM   | 16  | 32  | 32  | 16  | 16  | 16  | 16  | 16  | 32  | 16   |
| AF1  | 12  | 24  | 28  | 12  | 12  | 12  | 12  | 12  | 24  | 12   |
| AF2  | 12  | 24  | 24  | 12  | 8   | 16  | 8   | 16  | 32  | 8    |
| AF3  | 10  | 20  | 20  | 10  | 4   | 16  | 12  | 16  | 32  | 4    |
| AF4  | 8   | 16  | 16  | 8   | 16  | 16  | 16  | 16  | 32  | 0    |
| AF5  | 16  | 0   | 0   | 0   | 0   | 0   | 16  | 16  | 0   | 0    |
| AF6  | 16  | 0   | 0   | 16  | 0   | 0   | -16 | -16 | 0   | 0    |
| AF7  | -16 | 32  | -32 | 16  | -16 | -16 | 16  | 16  | 32  | -16  |
| AF8  | -6  | 4   | 4   | -6  | -4  | 0   | -4  | -8  | 0   | 0    |
| AF9  | 12  | 24  | 24  | 12  | 12  | 12  | 12  | 12  | 24  | 12   |
| AF10 | 8   | 16  | 16  | 8   | 4   | 4   | 0   | 0   | 8   | 4    |

Table S3. The relative energies (in meV) per 32 FU calculated for the 11 ordered spin states of BaYFeO$_4$ obtained for the 280, 38 and 6 K structures by DFT+U calculations

|  | 280 K | | 38 K | | 6 K | |
| --- | --- | --- | --- | --- | --- | --- |
| U$^{eff}$ | 4 eV | 5 eV | 4 eV | 5 eV | 4 eV | 5 eV |
| FM | 0.00 | 0.00 | 0.00 | 0.00 | 0.00 | 0.00 |
| AF1 | -752.03 | -636.97 | -766.00 | -648.30 | -744.78 | -628.97 |
| AF2 | -770.71 | -652.13 | -785.02 | -663.73 | -759.23 | -640.71 |
| AF3 | -1144.63 | -969.05 | -1165.92 | -986.34 | -1127.90 | -952.35 |
| AF4 | -1433.14 | -1218.23 | -1459.80 | -1240.00 | -1428.17 | -1210.03 |
| AF5 | -1642.82 | -1380.53 | -1672.93 | -1404.32 | -1627.40 | -1364.66 |
| AF6 | -1733.64 | -1451.29 | -1765.08 | -1475.94 | -1716.19 | -1433.63 |
| AF7 | -2922.76 | -2483.93 | -2979.54 | -2530.73 | -2892.52 | -2446.87 |
| AF8 | -3363.45 | -2849.21 | -3426.74 | -2900.67 | -3332.19 | -2812.91 |
| AF9 | -752.93 | -637.68 | -766.93 | -649.04 | -745.56 | -629.60 |
| AF10 | -1566.36 | -1321.93 | -1595.73 | -1345.58 | -1550.31 | -1304.31 |

Table S4. Values of $J_1 - J_{10}$ (in units of $k_BK$) obtained from DFT+U calculations with $U^{eff} = 4$ eV for the 280, 38 and 6 K structures.

(a) Intrachain spin exchange.

|       | J1     | J2    | J5    | J6   | J7   | J8   |
|-------|--------|-------|-------|------|------|------|
| 280 K | 153.05 | 86.10 | 9.91  | 3.00 | 2.65 | 3.09 |
| 38 K  | 155.93 | 87.59 | 10.10 | 3.10 | 2.69 | 3.18 |
| 6 K   | 150.74 | 85.44 | 7.94  | 5.15 | 2.54 | 3.89 |

(b) Interchain spin exchange.

|       | J3   | J4   | J9    | J10  |
|-------|------|------|-------|------|
| 280 K | 0.42 | 0.95 | 0.48  | 2.73 |
| 38 K  | 0.43 | 1.05 | 0.47  | 2.83 |
| 6 K   | 0.36 | 2.55 | -0.84 | 3.22 |

Table S5. The values of $J_1 - J_{10}$ (in units of $k_BK$) obtained from DFT+U calculations with $U^{eff} = 5$ eV for the 280, 38 and 6 K structures.

(a) Intrachain spin exchanges

|       | $J_1$  | $J_2$ | $J_5$ | $J_6$ | $J_7$ | $J_8$ |
|-------|--------|-------|-------|-------|-------|-------|
| 280 K | 131.50 | 73.04 | 7.86  | 2.00  | 2.12  | 2.12  |
| 38 K  | 133.93 | 74.24 | 8.00  | 2.05  | 2.15  | 2.17  |
| 6 K   | 129.17 | 72.39 | 6.26  | 3.48  | 2.02  | 2.65  |

(a) Interchain spin exchanges

|       | $J_3$ | $J_4$ | $J_9$  | $J_{10}$ |
|-------|-------|-------|--------|----------|
| 280 K | 0.33  | 0.27  | 0.61   | 2.06     |
| 38 K  | 0.34  | 0.33  | 0.62   | 2.13     |
| 6 K   | 0.30  | 1.34  | -0.29  | 2.43     |

Table S6. The relative energies $\Delta E(\|a)$, $\Delta E(\|b)$ and $\Delta E(\|c)$ in meV per 32 FUs of the three spin orientations $\|a$, $\|b$ and $\|c$, respectively, of BaYFeO$_4$ obtained by DFT+U+SOC calculations (with $U^{eff}$ = 4 eV) as well as the optimum spin-rotation angle $\theta_{min}$ defining the cycloidal magnetic structure of BaYFeO$_4$ predicted by the spin exchange constants $J_1$, $J_2$, $J_3$, $J_4$, $J_9$ and $J_{10}$ obtained by DFT+U (with $U^{eff}$ = 4 eV) calculations.

|  | 280 K | 38 K | 6 K |
|---|---|---|---|
| $\Delta E(\|a)$ | 5.13 | 5.42 | 4.76 |
| $\Delta E(\|b)$ | 0.00 | 0.00 | 0.00 |
| $\Delta E(\|c)$ | 1.18 | 1.62 | 1.00 |
| $\theta_{min}$ | 85.71° | 85.06° | 67.31° |

Table S7. The relative energies $\Delta E(\|a)$, $\Delta E(\|b)$ and $\Delta E(\|c)$ in meV per 32 FUs of the three spin orientations $\|a$, $\|b$ and $\|c$, respectively, of BaYFeO$_4$ obtained by DFT+U+SOC (with $U^{eff}$ = 5 eV) calculations as well as the optimum spin-rotation angle $\theta_{min}$ defining the cycloidal magnetic structure of BaYFeO$_4$ predicted by the spin exchange constants $J_1$, $J_2$, $J_3$, $J_4$, $J_9$ and $J_{10}$ obtained by DFT+U (with $U^{eff}$ = 5 eV).

|  | 280 K | 38 K | 6 K |
|---|---|---|---|
| $\Delta E(\|a)$ | 4.21 | 4.01 | 4.42 |
| $\Delta E(\|b)$ | 0.00 | 0.00 | 0.00 |
| $\Delta E(\|c)$ | 1.01 | 0.42 | 1.06 |
| $\theta_{min}$ | 92.02° | 91.49° | 75.04° |

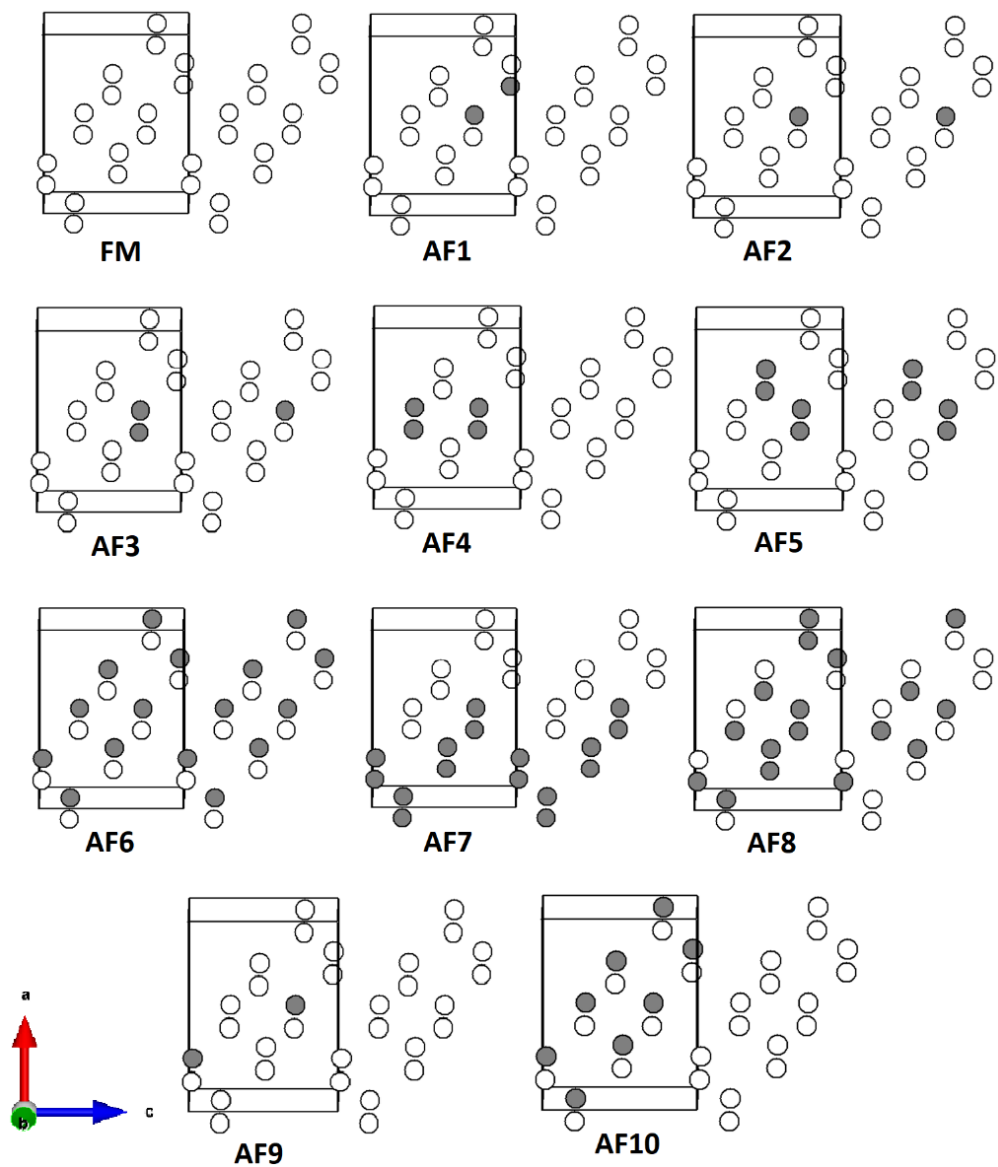

Figure S1. The 11 ordered spin states used to calculate $J_1 - J_{10}$.

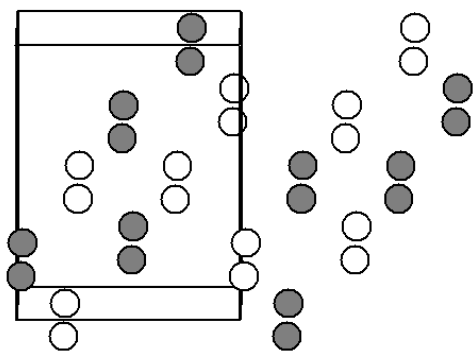

Figure S2. The AFM state used to determine the magnetic anisotropy of the $Fe^{3+}$ ions in BaYFeO$_4$.

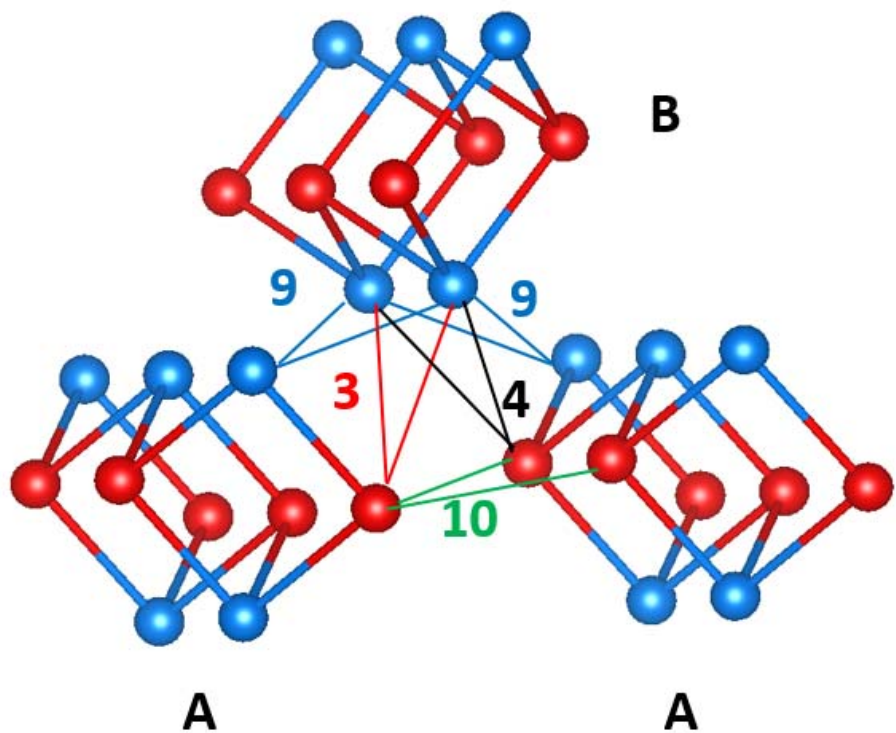

Figure S3. A perspective view of the interchain spin exchange paths $J_3$, $J_4$, $J_9$ and $J_{10}$. The intrachain spin exchanges $J_1$ and $J_2$ are also shown.

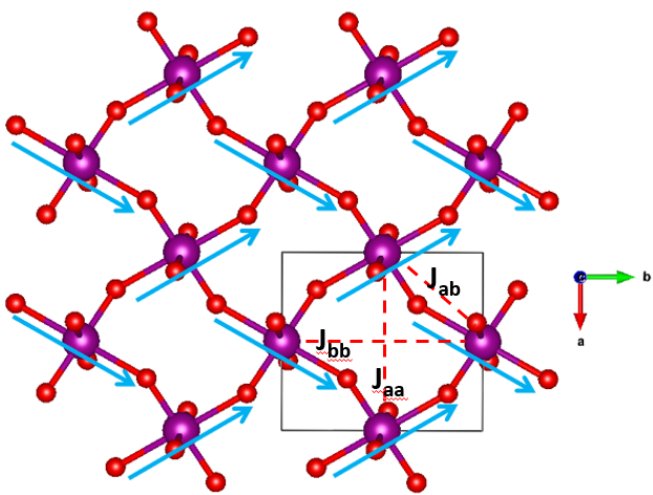

Figure S4. A layer of corner-sharing $MnO_6$ octahedra parallel to the ab-plane. The axially-elongated Mn-O bonds (indicated by blue arrows) lie in the ab-plane and are aligned more along the b-direction than along the a-direction.

The spin exchanges $J_{aa}$ and $J_{bb}$ are antiferromagnetic, but the spin exchange $J_{ab}$ is ferromagnetic. These spin exchanges, together with the preferred spin orientation of each $Mn^{3+}$ ion along the elongated Mn-O bond, makes the $Mn^{3+}$ spins lie in the bc-plane in the cycloidal magnetic structure of $TbMnO_3$ below 28 K.